# Measurement of Collective Dynamical Mass of Dirac Fermions in Graphene

Hosang Yoon[1], Carlos Forsythe[2], Lei Wang[3], Nikolaos Tombros[2], Kenji Watanabe[4], Takashi Taniguchi[4], James Hone[3], Philip Kim[2*], & Donhee Ham[1*]

[1] School of Engineering and Applied Sciences, Harvard University, Cambridge, MA 02138, USA.
[2] Department of Physics, Columbia University, New York, NY 10027, USA.
[3] Department of Mechanical Engineering, Columbia University, New York, NY 10027, USA.
[4] National Institute for Materials Science, Namiki 1-1, Tsukuba, Ibaraki 305-0044, Japan.
[*] Corresponding authors. E-mails: pk2015@columbia.edu, donhee@seas.harvard.edu.

**Individual electrons in graphene behave as massless quasiparticles[1-7]. In surprising twist, it is inferred from plasmonic investigations[8-11] that collectively excited graphene electrons must exhibit non-zero mass and its inertial acceleration is essential for graphene plasmonics. Despite such importance, this collective mass has defied direct unequivocal measurement. It may be directly measured by accelerating it with a time-varying voltage and quantifying the phase delay of the resulting current; this voltage-current phase relation would manifest as kinetic inductance, representing the collective inertia's reluctance to accelerate. However, at optical (infrared) frequencies phase measurement of current is generally difficult and at microwave frequencies the inertial phase delay has been buried under electron scattering[12-14]. Here we directly, precisely measure the kinetic inductance, thus, collective mass, by combining innovative device engineering that reduces electron scattering and delicate microwave phase measurements. Particularly, encapsulation of graphene between hexagonal-boron-nitride layers[15], one-dimensional edge contacts[16], and a proximate top gate configured as microwave ground[17,18] together enable resolving the inertial phase delay from the electron scattering. Beside the fundamental importance, the kinetic inductance demonstrated here to be orders-of-magnitude larger**

**than magnetic inductance can dramatically miniaturize radio-frequency integrated circuits. Moreover, its bias-dependency heralds a solid-state voltage-controlled inductor to complement the prevalent voltage-controlled capacitor.**

The collective excitation of massless fermions in graphene exhibits a non-zero mass. This fact is subsumed under the general theoretical framework of graphene plasmonics[8] [Supplemental Information (SI)], yet it can be simply seen as follows. Let electrons in graphene (width $W$, unit length) be subjected to an electric field along the length with a voltage difference $V$ across the length. The resulting translation of the Fermi disk in two-dimensional $k$-space by $\Delta k \ll k_F$ (from disk **A** to **B**, Figs. 1a,b) yields a per-unit-length collective momentum, $P = n_0 W \hbar \Delta k$ ($n_0$: electron density). The corresponding per-unit-length collective kinetic energy $E$ is obtained by subtracting the sum of single electron energies $\varepsilon = \hbar v_F k$ for disk **A** from that for disk **B**. Since $E$ is minimal at $\Delta k = 0$, we must have $E \propto (\Delta k)^2 \propto P^2$ for small $\Delta k$ (Fig. 1c). In fact, calculation to the lowest order of $\Delta k$ (SI) shows $E = W\varepsilon_F/2\pi \times (\Delta k)^2 = P^2/2M$, where $M = \pi W n_0^2 \hbar^2/\varepsilon_F$ is the collective mass per unit length. This remarkable emergence of non-zero collective mass with the quadratic $E$-$\Delta k$ relation from massless individual electrons with the linear $\varepsilon$-$k$ relation sharply contrasts with the case of typical conductors with quadratic single electron energy dispersion $\varepsilon = \hbar^2 k^2/(2m^*)$, where the collective mass is simply the sum of the non-zero individual electron masses $m^*$. Incidentally, we note that the collective mass of graphene electrons is quantitatively related to an insightfully defined theoretical entity called 'plasmon mass' in graphene[9,11,19]; the former, which we set out to measure in this work, is an observable physical reality that proves the existence of the latter beyond a theoretical model.

The collective current $I$ associated with the Fermi disk shift—that is, the inertial acceleration of the collective mass $M$—has an inductive phase relationship to the applied voltage $V$ that causes the acceleration, where the associated inductance is kinetic inductance. The kinetic inductance can be evaluated by noting $E \propto I^2$, given $I \propto \Delta k$ for small $\Delta k$ and $E \propto (\Delta k)^2$; by analogy to magnetic inductance, this energy can be then expressed as $E = L_k I^2/2$, where

$$L_k = \pi\hbar^2/(We^2\varepsilon_F) = \hbar\sqrt{\pi}/(We^2 v_F) \times 1/\sqrt{n_0} \qquad (1)$$

is the per-unit-length kinetic inductance (SI). The same underlying physics, namely quadratic dependence of $E$ on $\Delta k$, gives rise to both $M$ and $L_k$, which are thus intimately related by $M = (e^2 n_0^2 W^2) \times L_k$; in fact, the kinetic inductance represents the 'inertial' reluctance of the collective current to change. Incidentally, the peculiar $L_k \sim 1/\sqrt{n_0}$ dependence arising from graphene's linear single-electron $\varepsilon$-$k$ dispersion contrasts the $L_k \sim 1/n_0$ dependence[17,18,20] in typical conductors with quadratic single-electron $\varepsilon$-$k$ dispersion.

To weigh $M$, we directly measure $L_k$ essentially by interrogating the voltage-current phase relation in graphene via microwave transport experiment, where graphene acts as a lossy transmission line[17,18,20-22] (Fig. 1d). This entails the per-unit-length kinetic inductance $L_k$ modeling local collective mass and per-unit-length geometric capacitance $C$ terminated to ground specific to device configuration. The magnetic inductance, which is orders of magnitude smaller[17,22] than $L_k$ expected in graphene (SI), and quantum capacitance[23], whose effect is far weaker than that of $C$ in our device geometry to be discussed, are both ignored. The per-unit-length resistance $R$ models electron scattering.

While the graphene kinetic inductance can be implied from the plasmonic theory[8] and has been considered in explicit theories[19,24,25], its direct measurement has been evasive. In far-infrared intensity transmission spectroscopies, kinetic inductance can be indirectly inferred from the fitting parameter called Drude weight[9,11,19], but as these experiments do not measure the phase progression of the collective current, they do not unambiguously prove the existence of the collective mass and its inertial acceleration. At microwave frequencies, while $L_k$ can in principle be directly measured by probing the voltage-current phase relation, experimental attempts[12-14] have proven unfruitful because $R$ is far larger than the inductive impedance $i\omega L_k$ at microwave frequencies (*i.e.*, the kinetic inductor's quality factor $Q = \omega L_k/R$ is far smaller than 1) even in reasonably high-mobility graphene, although $L_k$ is far larger than magnetic inductance.

In our microwave measurements of $L_k$, we help overcome this difficulty by reducing electron scattering, thus, $R$, as much as possible. In particular, we encapsulate exfoliated graphene between two hexagonal boron nitride (h-BN) layers (Fig. 2) by a polymer-free assembly method[15,16], which greatly reduces electron scattering by disorder. To reduce electron-phonon scattering within graphene, the device is cooled to 30 K. Furthermore, to reduce additional electron scattering in the contact regions at both ends of graphene, we make one-dimensional, edge-only contacts to the graphene by etching the stack of h-BN and graphene into a desired shape ($W = 7.5$ μm; length $l = 19.0$ μm) and depositing metal onto the side edges[16] (Figs. 2a-c). Graphene is connected to the signal (S) lines of on-chip coplanar electromagnetic waveguides (CPWs) to the left and right via the abovementioned edge-only contacts, and is placed under a top gate merged with the ground (G) lines of the CPWs.

We first measure the DC resistance between the S lines using a lock-in technique, with the graphene and top gate kept at the same DC potential. The total device resistance $R_{dev}$, including the effect of both $R$ and contact resistance, is measured as a function of the back gate potential $V_b$ that sets the carrier density $n_0$ (Fig. 2d). At 30 K, it shows a charge neutrality at $V_{b,0} = -0.5$ V and excellent performance in the electron-doped region ($V_b > V_{b,0}$), which is fit well by the widely-adopted conductivity formula[15,26] $\sigma^{-1} = (n_0 e \mu_C)^{-1} + \rho_s$, with $\mu_C = 390,000$ cm$^2$/Vs representing the $n_0$-independent mobility due to long-range scattering, and $\rho_s = 80$ Ω representing the short range scattering. We note that this is only a lower bound of the actual graphene mobility, because the estimation of conductivity in this two-probe measurement includes the contact effects. The hole-doped region ($V_b < V_{b,0}$) shows similar results, but with slightly lower $\mu_C = 320,000$ cm$^2$/Vs and higher $\rho_s = 110$ Ω due to the contact characteristics[16]. The room-temperature measurement shown for comparison (which exhibits a slightly shifted neutrality at $V_{b,0} = -0.9$ V) has a ~4 times smaller $\mu_C$ of 110,000 cm$^2$/Vs in the electron-doped regime, which still is an excellent number. This high $\mu_C$ at room temperature confirms the reduction in electron scattering by the h-BN encapsulation[15,16], and its 4-fold increase at 30 K confirms the further scattering reduction at the low temperature.

To measure $L_k$, a vector network analyzer launches microwaves (10-50 GHz) onto the CPWs, and records the amplitude and phase response of their transmission ($s_{21}$, $s_{12}$) and reflection ($s_{11}$, $s_{22}$) (Fig. 3a). The network analyzer connects to graphene via cables, probes, and the CPWs, whose phase delay and loss are calibrated out. The direct parasitic coupling between the left and right CPWs/probes bypassing the graphene channel is separately measured and de-embedded. We extract $L_k$ from the resulting $s$-parameters. This extraction, however, poses a stiff challenge for moderate-mobility graphene with $Q \ll 1$, which can be appreciated from the expression for the per-unit-length phase delay $\phi$ through the graphene transmission line (Fig. 1d); $\phi \approx (\omega RC/2)^{1/2} + (\omega^3/8)^{1/2}(C/R)^{1/2} \times L_k \equiv \phi_1 + \phi_2$, with only $\phi_2$ containing $L_k$ (SI), capturing the phase delay due to the collective mass acceleration. The ratio $\phi_2/\phi_1 = \omega L_k/R/2 = Q/2$; with $Q \ll$ 1, extraction of $L_k$ is challenging because $\phi_2$ is entirely swamped by $\phi_1$, which typically renders $\phi_2$ itself miniscule below the unavoidable phase measurement error—which we denote as $\phi_e$—caused by imperfect calibration and non-ideal parasitic signal de-embedding.

To enable $L_k$ extraction from the measured *s*-parameters, we first reduce $R$ via the aforementioned h-BN encapsulation of graphene and 30-K operation, which amplifies $\phi_2$ and attenuates $\phi_1$ with improved $Q = 2\phi_2/\phi_1$. This crucial improvement alone, however, is insufficient with the improved $Q$ still smaller than 1. A second improvement is to enhance $C$ by proximate top gating. Although increasing $C$ does not change $Q = 2\phi_2/\phi_1$, it further increases $\phi_2$ to ensure $\phi_2 > \phi_e$. Importantly, these improvements also make $\phi_2$ more sensitive to $L_k$ variation, as seen from the factor $(C/R)^{1/2}$ in $\phi_2$, thereby increasing the accuracy of $L_k$ extraction (SI). To estimate the enhanced value of $C$ in our device, we note that the proximate top gate merged with the CPWs' G lines (Figs. 2a-c) serves as a well-defined microwave ground[17,18] with per-unit-length capacitance $C_g$ between graphene and this grounded top gate. In contrast, the silicon back gate *un*tapped to the G lines 'floats' in microwave signaling, largely because its connection to the DC bias line exhibits a very large inductive impedance and also because the silicon has a high resistivity. Therefore, the per-unit-length capacitance $C_b$ between graphene and the back gate is irrelevant for microwave signaling, and $C = C_g$. As 44-nm thick top h-BN ($\kappa \approx 7$ [27,28]) and ~150-nm thick hydrogen silsesquioxane (HSQ; $\kappa \approx 2.8$~3.0 [29]) lie between graphene and the grounded top gate, $C_g/W$ is estimated to be 0.15 fF/μm$^2$, which is far larger than the capacitance of ungated graphene[21]. Incidentally, we attribute the inability to observe[12,13] $L_k$ or its spurious measurement[14] in prior works to their larger $R$ with graphene on SiO$_2$ and no proximate gate configured as a microwave ground.

Figure 3b is a color map of the phase and amplitude of transmission ($s_{21}$) and reflection ($s_{11}$) parameters measured at 30 K as functions of $V_b$ (thus $n_0$) and frequency. The $s_{21}$ [$s_{11}$] amplitude exhibits a sharp drop [peak] near $V_{b,0}$ = -0.5 V. Fig. 3c shows $s_{21}$ at three select $V_b$ values in the electron-doped region to show that our device $s$-parameters are amenable to $L_k$ extraction. Were it not for the $R$-reduction and $C$-enhancement, the measured $s_{21}$ phase and its portion contributed by $L_k$—which are intimately related to $\phi$ and $\phi_2$, respectively—would exhibit far smaller absolute values as well as far smaller differences with the variation of $V_b$ (thus with the corresponding variation of $L_k$), hampering $L_k$ extraction (SI).

To determine $L_k$ from the measured $s$-parameters for each bias, we use the microwave optimization method[17]; we add contact models to both sides of the transmission line model (Fig. 1d with $C = C_g$), and alter the component values (*e.g.*, $L_k$, $C_g$, $R$, and contact resistance) until the $s$-parameters calculated from the model best fit the measured $s$-parameters across the frequency range in the least-square method (SI). In this way, we determine $L_k$ and other component values at each $V_b$. This method's reliability is based on the model's physicality and the fact that the limited number of model components must reproduce the vastly larger number of measured $s$-parameters over the frequency range. Its cogency will be checked ultimately by the consistency amongst the extracted values and other measured parameters, and with the physical theory. The same experiment repeated on a completely different device led to almost identical results (SI), further attesting to the reliability of this approach.

Figures 4a-c display $L_k$, $C_g$, and $R$ so determined for each $V_b$ at 30 K and 296 K. We first focus on the 30-K results in the electron-doped region ($V_b > V_{b,0}$) that showed the best DC characteristics (Fig. 2d), in particular in the region away from $V_{b,0}$ (unshaded region in Fig. 4). As expected, the extracted $C_g/W$ stays nearly constant with negligible variation from quantum

capacitance effect[23], and its value of ~0.15 fF/μm$^2$ is consistent with the value roughly estimated earlier. Also, the extracted $L_k$ closely follows the theoretical curve given by equation (1) with $v_F$ = 10$^6$ m/s. The slight discrepancy between the observed and theoretical $L_k$ in this region is attributed dominantly to imperfect calibration and parasitic-signal de-embedding, but also potentially to variations of $v_F$ due to dielectric screening and impurities[28], and/or electron-electron interaction effects[19]. Further confirming the consistency of the technique, $R_{dev}$ extracted from the *s*-parameters agrees well with $R_{dev}$ measured at DC (Fig. 4c). Most importantly, from the measured $L_k$, we obtain the per-unit-length collective mass, $M = L_k \times e^2 n_0^2 W^2$, or *operationally* defined collective mass *per electron* that Ref. 19 theorizes as 'plasmon mass,' $m^*_c$ = $M/(Wn_0) = L_k \times e^2 n_0 W$, which closely follows the theoretical prediction (Fig. 4a, inset); $m^*_c$ is a few percent of $m_0 = 9.1 \times 10^{-31}$ kg.

Near the charge neutrality point or in the hole-doped region ($V_b < V_{b,0}$) (shaded region, Fig. 4), the extracted values of $L_k$, $C$, and $m^*_c$ at 30 K exhibit more appreciable deviation from theory. The discrepancy near the charge neutrality is readily understood, because transmission amplitude is significantly smaller due to the sharply reduced $n_0$ (Fig. 3b). In this region, the raw transmission *s*-parameters before removing the graphene-bypassing parasitic signal are dominated by the parasitic signal itself, making the parasitic-signal-de-embedded *s*-parameters highly distorted. The best-optimized model *s*-parameters then still poorly fit the distorted s-parameters, for our model does not take into account the distortion effect (SI). The discrepancy in the hole-doped region is similarly explained, as the measured signal is distorted (SI) in a way that cannot be fully captured by the model in use. This distortion can be traced back to the asymmetric behavior caused by work function mismatch in our edge contacts, where the contact between the metal and hole-doped graphene has been demonstrated to exhibit non-ideal behaviors[16] that are difficult to capture with a passive linear model (SI).

Back in the higher-fidelity electron-doped region (unshaded region in Fig. 4), the data at 296 K result in more appreciable deviation from theory, due to the ~4× decrease in mobility (~4× increase in $R$), which reduces all of $\phi_2/\phi_1$, $\phi_2$, $(C/R)^{1/2}$, and transmission amplitude. This highlights the challenge in measurements of sub-unit $Q$ devices. Nonetheless, while not as quantitatively accurate as the 30-K data, the 296-K data still present a firm direct proof of $L_k$ and collective dynamical mass, made possible by the h-BN graphene interface and the proximate gating. Thus even the 296-K data represent a significant leap from the prior works that have only failed to observe the kinetic inductance[12-14].

Beside its fundamental importance for graphene electrodynamics and plasmonics, our work may offer exciting technological vistas. The graphene kinetic inductance as a manifestation of the collective inertia effect is orders of magnitude larger than the magnetic inductance at similar dimensions (SI), and thus can be used in the future to substantially miniaturize inductors, as it allows one to obtain the same inductance value in orders of magnitude smaller area. Radio-frequency integrated circuits, such as resonators, filters, oscillators, and amplifiers, prevalent in communication and computing systems, suffer from large chip areas due to magnetic inductors. Thus these high frequency applications may benefit greatly from harnessing the kinetic inductance of graphene revealed in this work. Better room temperature scalability and facile tunability as compared to traditional kinetic inductors from superconductors and semiconductor two-dimensional gases also bode well in this direction as the mobility of graphene continues to improve. Furthermore, the bias-dependency of graphene kinetic inductance renders graphene a natural voltage-controlled tunable inductor as a counterpart to the prevalent voltage-controlled semiconductor capacitor.

**Methods Summary**

We fabricated h-BN encapsulated graphene by mechanical exfoliation and polymer-free mechanical transfer of h-BN single crystals and graphene through optical alignment[16]. High resistivity (> 5000 Ω cm) silicon wafers coated with 285-nm thick thermal oxide were used as the substrate to minimize high-frequency substrate losses. Optical differentiation and Raman spectroscopy were used to confirm that graphene is single layered. Contacts[16] and waveguides were created by thermal evaporation of Cr/Pd/Au (1/10/300 nm) with dimensions defined by electron beam lithography and inductively coupled plasma etching.

Measurements took place in a Lake Shore Cryotronics cryogenic probe station at feedback-controlled temperatures in the dark. DC resistance measurements were performed using a Stanford Research Systems SR830 lock-in amplifier and a DL Instruments 1211 current preamplifier. Microwave *s*-parameter measurements were performed using an Agilent E8364A vector network analyzer, where the calibration was performed using the NIST-style multiline TRL technique[30] at each temperature just before the measurement. The parasitic coupling bypassing the graphene device was measured on a separate device with the identical CPW structures but with no h-BN encapsulated graphene, and was then de-embedded from the measured *s*-parameters of the main device[22].

The design of the CPWs was performed using a Sonnet frequency-domain electromagnetic field solver. The CPW dimensions were chosen to match the 50-Ω characteristic impedance of the network analyzer, cables, and probes[22].

**Supplementary Information** is linked to the online version of the paper at www.nature.com/nature.

**Acknowledgments** D.H. and H.Y. are grateful for the support of this research by Samsung Advanced Institute of Technology and its Global Research Opportunity program under contract no. A18960 and by the Air Force Office of Scientific Research under contracts no. FA 9550-09-1-0369 and no. FA 9550-08-1-0254. P.K. acknowledges Nano Material Technology Development Program through the National Research Foundation of Korea (NRF) funded by the Ministry of Science, ICT and Future Planning (2012M3A7B4049966). J.H. and L.W. acknowledge support from NSF DMR-1124894, and the Office of Naval Research under Award N000141310662. C.F. acknowledges support from IGERT program. The device fabrication was performed in part at the Center for Nanoscale Systems at Harvard University.

**Author Contributions** H.Y., P.K., and D.H. conceived the project. K.W. and T.T. fabricated the h-BN. H.Y., C.F., L.W., N.T., and J.H. fabricated the stacked layers of h-BN, graphene, and h-BN. H.Y. designed the device. H.Y. and C.F. fabricated the device. H.Y. performed the experiments. H.Y., P.K., and D.H. analyzed the data. H.Y., P.K., and D.H. wrote the paper. All authors discussed the results and reviewed the manuscript.

**Author Information** Reprints and permissions information is available at www.nature.com/reprints. The authors declare no competing financial interest. Correspondence and requests for materials should be addressed to P.K. (pk2015@columbia.edu) and D.H. (donhee@seas.harvard.edu).

**Figure 1 | Collective electrodynamics of graphene electrons. a**, Collective motion of graphene electrons subjected to an electric field can be represented as a translation of Fermi disk in $k$-space. **b**, Alternative representation of the collective electron motion in the $\varepsilon$-$k$ space, in conjunction with the massless single electron energy dispersion $\varepsilon = \hbar v_F k$ near the Dirac point. **c**, The per-unit-length collective kinetic energy $E$ exhibits quadratic dependency on the per-unit-length collective momentum $P = n_0 W \hbar \Delta k$. The curvature is inversely proportional to the per-unit-length collective dynamical mass, $M$. **d**, Graphene as a lossy transmission line.

**Figure 2 | Device description and DC measurements. a**, Optical image of the h-BN/graphene/h-BN layered structure before etching (top-left), after etching (top-right), and after depositing the CPWs (bottom). **b**, False colored scanning electron micrograph of the central region of the device that contains the layered structure under the top gate. **c**, Schematic diagram of h-BN encapsulated graphene device with the front face corresponding to the vertical cut through the dotted line in **b**. **d**, Total device resistance $R_{dev}$, including both in-graphene electron scattering effect $R$ and contact resistance, measured at 30 K and 296 K with $V_b$ varied while graphene and the top gate are kept at the same DC potential. (inset: corresponding plot of $(R_{dev}/(l/W))^{-1}$, a conductivity estimate including contact effects; $n_0 = C_b/W \times (V_b - V_{b,0})/e$ with $C_b/W$ = 0.12 fF/μm$^2$ and $V_{b,0}$ = -0.5 V). Red solid curves are fits to $\sigma^{-1} = (n_0 e \mu_C)^{-1} + \rho_s$.

**Figure 3 | Microwave $s$-parameter measurements. a**, Schematic diagram of the measurement setup. The $s$-parameters shown are after calibrating out the delay and loss of the cables, probes, and on-chip CPWs, and also after de-embedding the parasitic coupling bypassing graphene. **b**,

Phase (insets: amplitude) of the measured transmission ($s_{21}$; left) and reflection ($s_{11}$; right) parameters after the calibration and de-embedding at 30 K. The *s*-parameters with excitation from the opposite side ($s_{12}$ and $s_{22}$; not shown here) look almost identical to $s_{21}$ and $s_{11}$. **c,** Select data from **b**, specifically, transmission phase ($\angle s_{21}$; solid curves) and amplitude ($|s_{21}|$; dashed curves) at three representative bias values $V_b = 1$, 4, and 20 V.

**Figure 4 | Extracted graphene kinetic inductance and collective electron mass.** Kinetic inductance per square, $L_k W$ (**a**), graphene to top-gate capacitance per unit area, $C_g/W$ (**b**), total device resistance, $R_{dev}$ (**c**), and collective dynamical mass per electron, $m^*_c$ (**d**), extracted from the measured *s*-parameters for various $V_b$ at 30 K and 296 K. The solid curves in **a**, **b**, and **d** represent theoretical predictions. The solid curve in **c** is $R_{dev}$ measured at DC (Fig. 2d). The shaded areas indicate bias regions where the extraction was less reliable (see text).

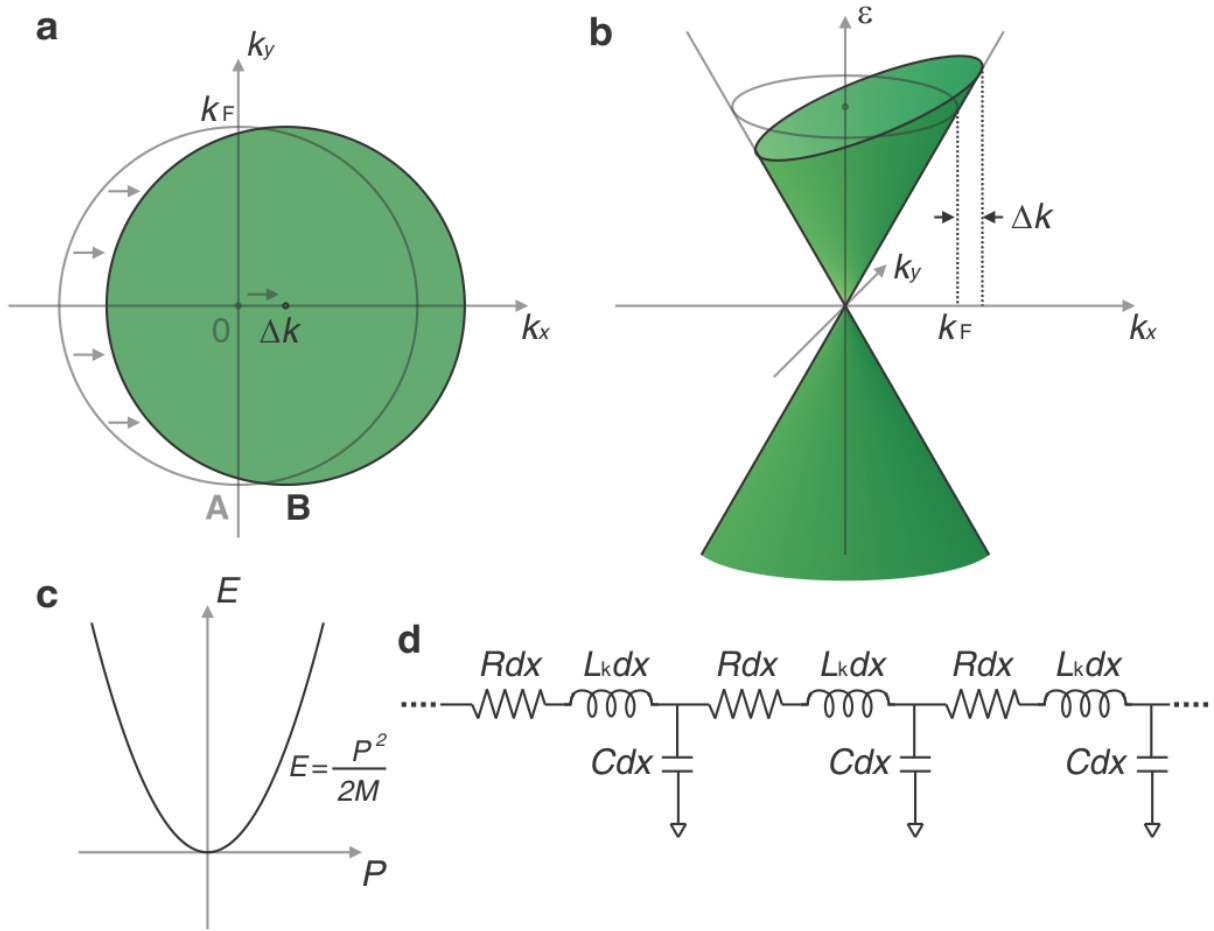

Figure 1

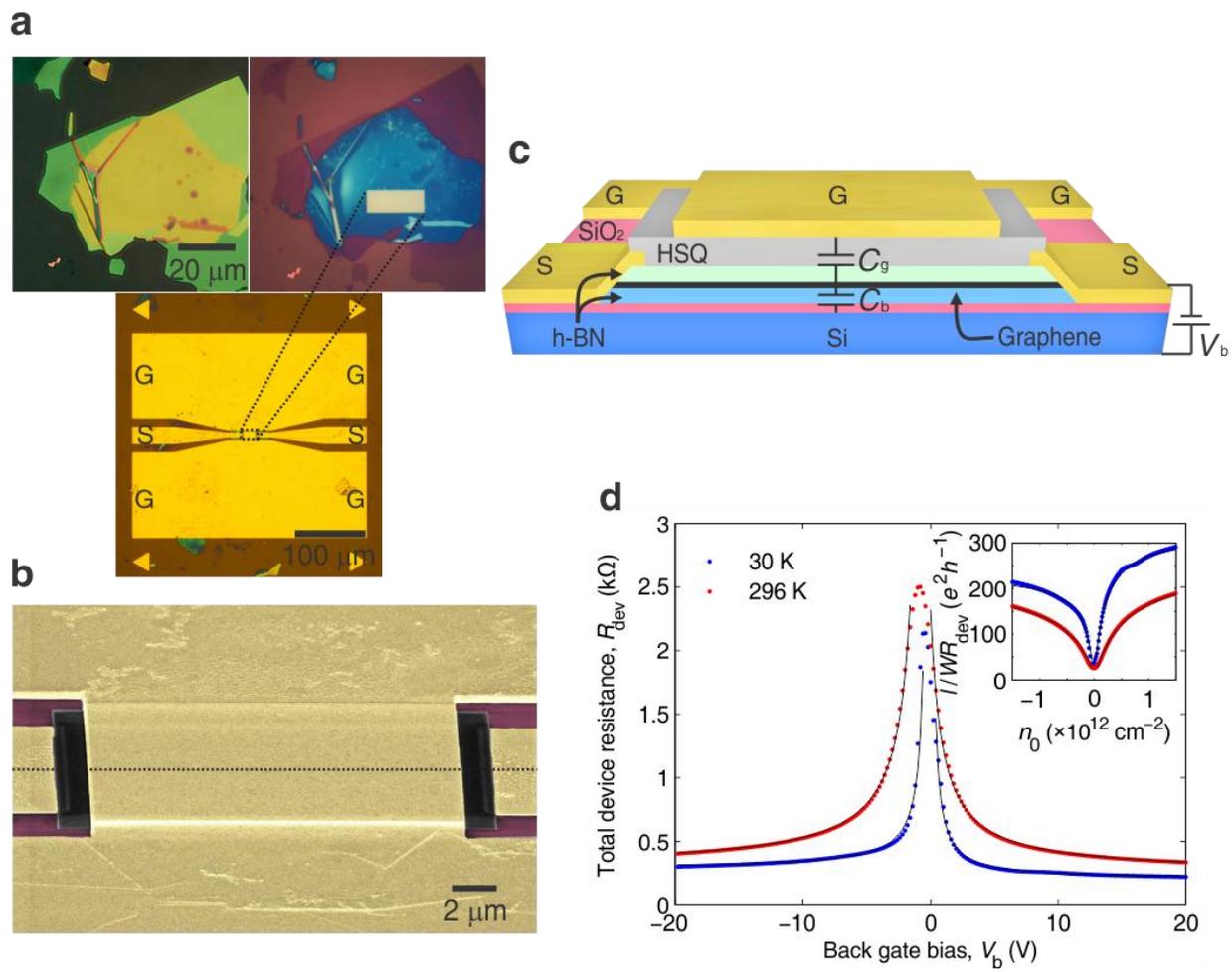

Figure 2

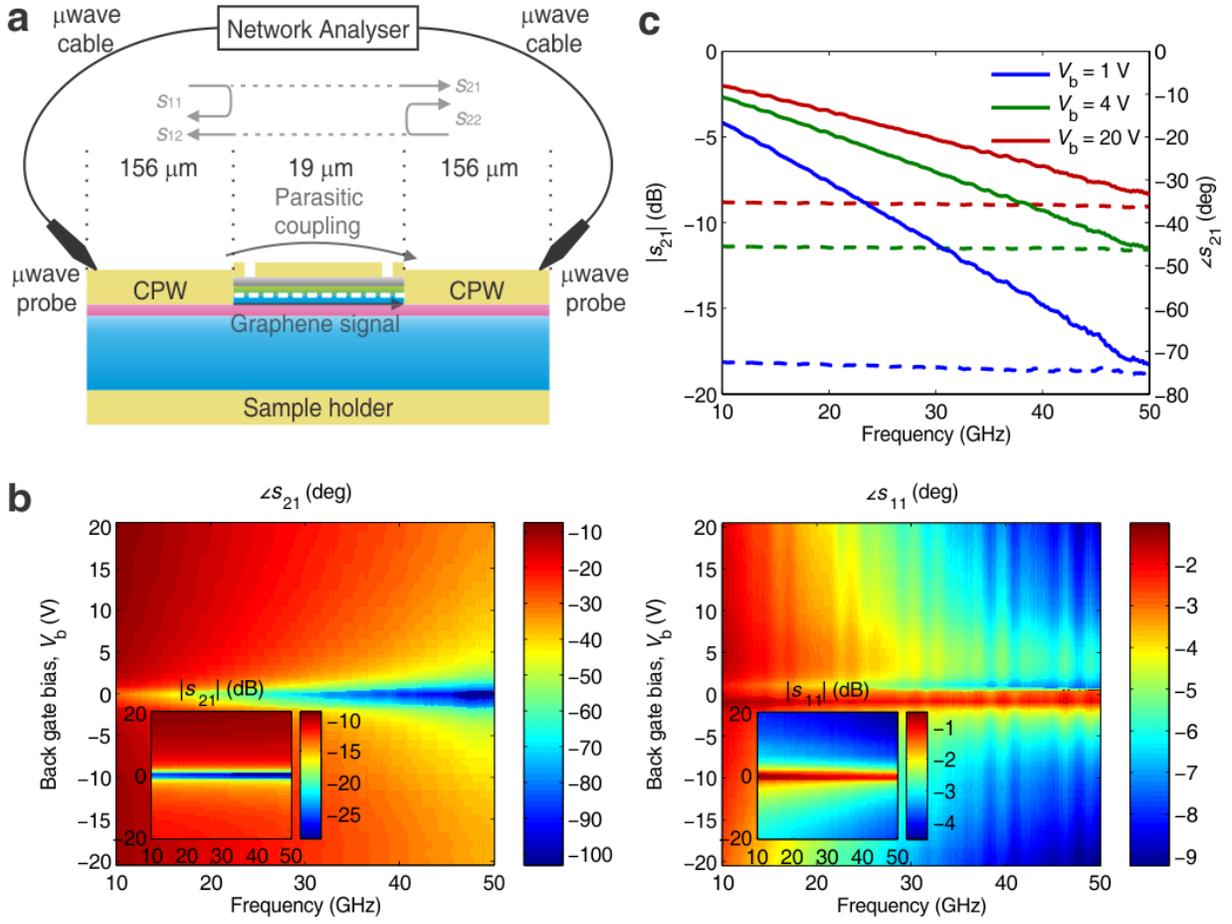

Figure 3

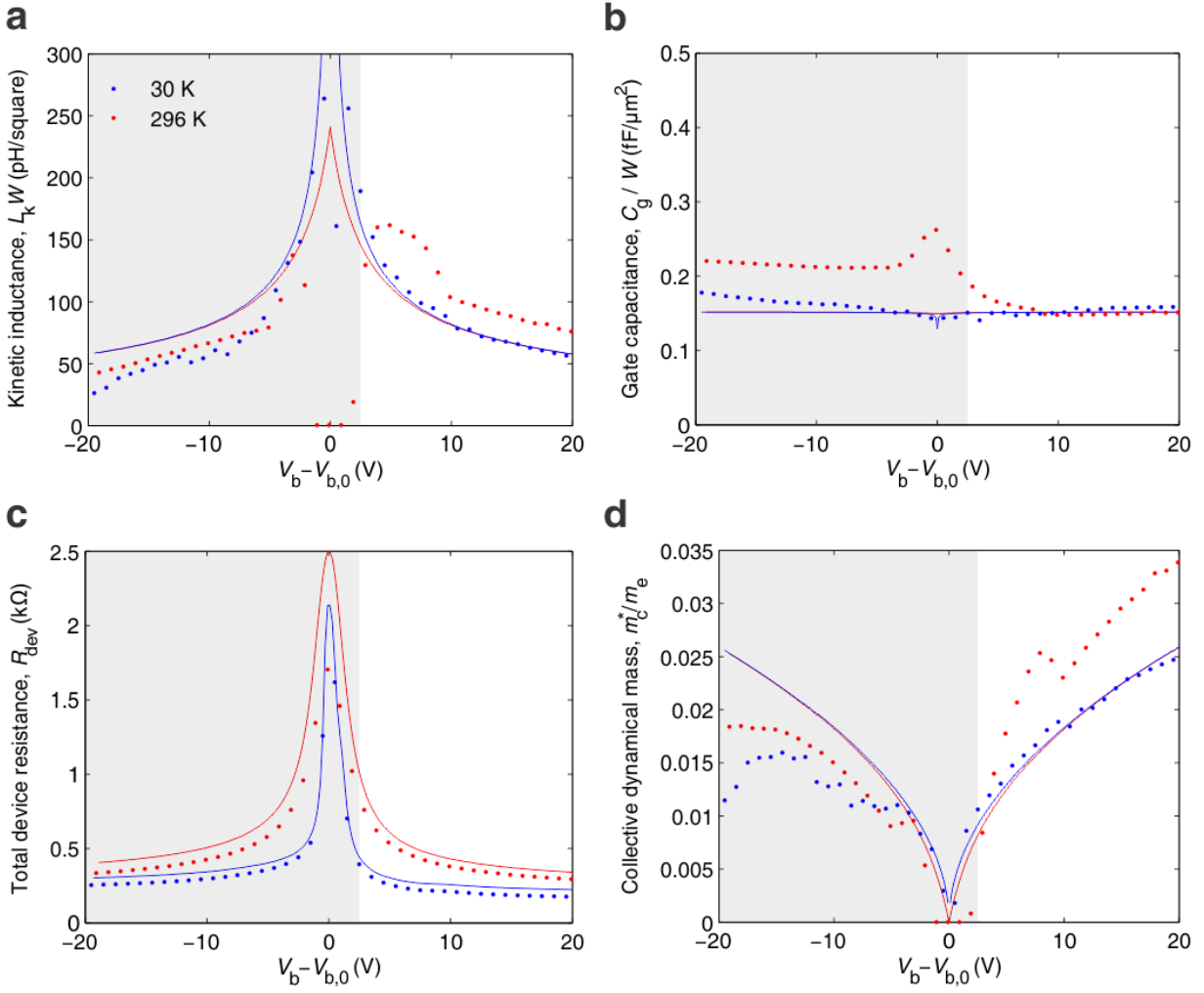

Figure 4